\documentclass[
 reprint,
superscriptaddress,
amsmath,amssymb,
]{revtex4-2}
\usepackage{float} % to allow H in table positioning 
\usepackage{graphicx}% Include Figure files
\usepackage{dcolumn}% Align table columns on decimal point
\usepackage{bm}% bold math
\usepackage{verbatim}

\usepackage{caption}
\usepackage{subcaption}
\graphicspath{ {./images/} }

\usepackage{array}
\usepackage{hyperref}  

\usepackage[T1]{fontenc}
\DeclareMathAlphabet{\mathpzc}{T1}{pzc}{m}{it}

\makeatletter
\def\blfootnote{\gdef\@thefnmark{}\@footnotetext}
\makeatother

\begin{document}

\preprint{}

\title{Delay model for the dynamics of information units in the digital environment} 

\author{Sebastián Pinto}
\email{spinto@df.uba.ar}
\affiliation{Universidad de Buenos Aires, Facultad de Ciencias Exactas y Naturales, Departamento de Física, Ciudad Universitaria, 1428 Buenos Aires, Argentina.}
\affiliation{CONICET - Universidad de Buenos Aires, Instituto de Física Interdisciplinaria y Aplicada (INFINA), Ciudad Universitaria, 1428 Buenos Aires, Argentina.}

\author{Alejandro Pardo Pintos}
\affiliation{Universidad de Buenos Aires, Facultad de Ciencias Exactas y Naturales, Departamento de Física, Ciudad Universitaria, 1428 Buenos Aires, Argentina.}
\affiliation{CONICET - Universidad de Buenos Aires, Instituto de Física Interdisciplinaria y Aplicada (INFINA), Ciudad Universitaria, 1428 Buenos Aires, Argentina.}

\author{Pablo Balenzuela}
\affiliation{Universidad de Buenos Aires, Facultad de Ciencias Exactas y Naturales, Departamento de Física, Ciudad Universitaria, 1428 Buenos Aires, Argentina.}
\affiliation{CONICET - Universidad de Buenos Aires, Instituto de Física Interdisciplinaria y Aplicada (INFINA), Ciudad Universitaria, 1428 Buenos Aires, Argentina.}

\author{Marcos Trevisan}
\affiliation{Universidad de Buenos Aires, Facultad de Ciencias Exactas y Naturales, Departamento de Física, Ciudad Universitaria, 1428 Buenos Aires, Argentina.}
\affiliation{CONICET - Universidad de Buenos Aires, Instituto de Física Interdisciplinaria y Aplicada (INFINA), Ciudad Universitaria, 1428 Buenos Aires, Argentina.}

\date{\today}% It is always \today, today,

\begin{abstract}
The digital revolution has transformed the exchange of information between people, blurring the traditional roles of sources and recipients. In this study, we explore the influence of this bidirectional feedback using a publicly available database of quotes, which act as distinct units of information flowing through the digital environment with minimal distortion. Our analysis highlights how the volume of these units decays with time and exhibits regular rebounds of varying intensities in media and blogs. To interpret these phenomena, we introduce a minimal model focused on delayed feedback between sources and recipients. This model not only successfully fits the variety of observed patterns but also elucidates the underlying dynamics of information exchange in the digital environment. Data fitting reveals that the mean attention to an information unit decays within approximately 13 hours in the media and within 2 hours in the blogs, with rebounds typically occurring between 1 and 4 days after the initial dissemination. Moreover, our model uncovers a functional relationship between the rate of information flow and the decay of public attention, suggesting a simplification in the mechanisms of information exchange in digital media. Although further research is required to generalize these findings fully, our results demonstrate that even a bare-bones model can capture the essential mechanisms of information dynamics in the digital environment.
\end{abstract}


%\tableofcontents
\maketitle

\section*{Introduction}

% Intro general
To effectively understand the mechanisms that drive the flow of information in the digital environment, there is a pressing need to study how attention to novel items propagates and fades among large populations \cite{Wu2007}.  The dynamics of public attention have been approached from different methodologies, including dynamical systems, epidemiological models, stochastic processes, auto-regressive models and agent-based models among others \cite{lorenz2019accelerating, towers2015mass, rizoiu2017expecting, teng2017dynamic, baumann2020modeling}.

Specifically, dynamical systems offer a minimalist approach aimed at capturing the fundamental mechanisms underlying the dynamics of cultural objects. For example, Candia and collaborators analyze the decay of attention to scientific publications, songs, movies and biographies using a two-dimensional mathematical model for the interaction between communicative and cultural memory \cite{candia2019universal}. Another approach implements a Lotka-Volterra model to describe the effect of imitation, competition and decay of interest that successfully reproduces the observed data \cite{lorenz2019accelerating}. Here we capitalize on a work from Leskovec and collaborators \cite{leskovec2009meme} that develop a framework for tracking short, distinctive phrases that travel relatively intact through on-line text. The authors set up a scalable algorithm for clustering textual variants of quotes that they call `memes'. These memes act as units of information that travel undisturbed along the information flow, providing a coherent representation of the news cycle. The result is a publicly available database that tracked 1.6 million mainstream media sites and blogs in England over a period of three months of 2008, for a total of circa 90 million articles. The authors also proposed a simple dynamical model for the imitation and recency (new threads are favored over older ones) that qualitatively reproduces the experimental data. 

In this work we re-analyze the data provided in this database \cite{leskovec2009meme} and identify a set of persistent temporal patterns that can be understood in terms of a feedback between sources and recipients. With this novel interpretation, in which media and blogs express a general model of sources and recipients of information, we set up a minimalist dynamical model that successfully fits the observed patterns. The analysis also allows disclosing the relationship between rates of transfer and allows estimating typical reaction times.

\section*{Data}

\begin{figure*}[htbp]
    \centering
    \includegraphics[width = \textwidth]{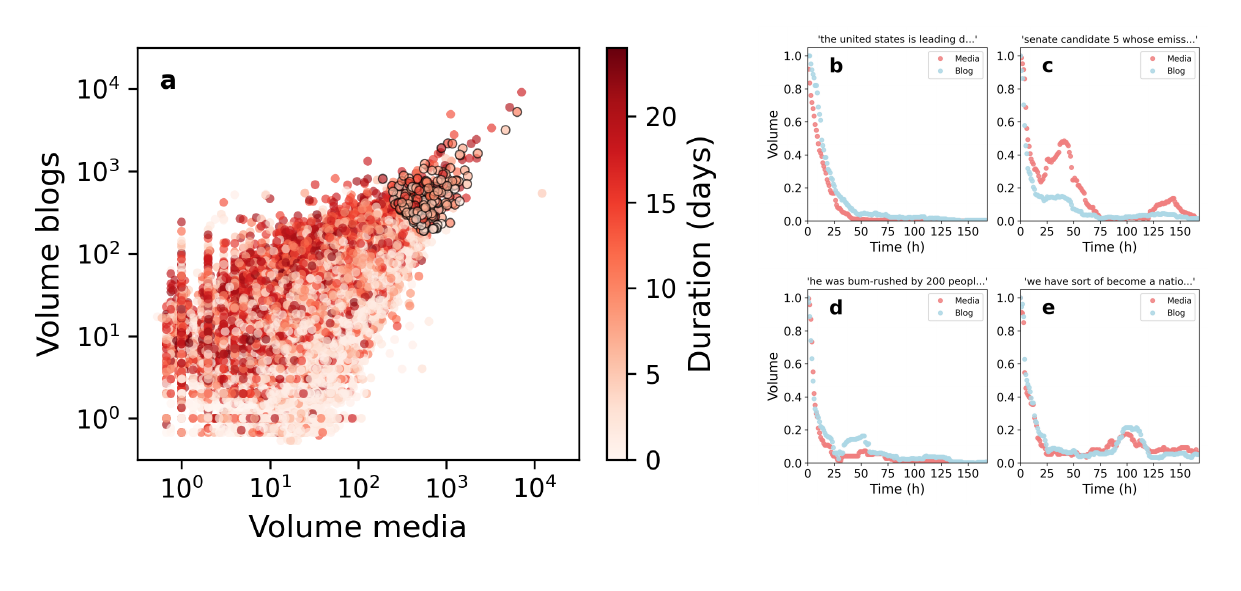}
\caption{\textbf{Volume of mentions produced by the information units and patterns of temporal evolution.} \textbf{a}. We estimate the volume of quotes by summing the mentions in a 3-day period around the peak. Most of the quotes receive very few mentions and survive in the system for a short time of around 2 days. However, there is a subset of approximately 200 quotes circled in black that survive in the system for more than a week, and produce time series with good signal to noise ratio. The dynamics observed for the volume of quotes can be organized in four families: \textbf{b.} exponential decays, which constitutes around the 36\% of the selected subset; \textbf{c.} damped rebounds which are prominent in media (21\% of the subset); \textbf{d.} damped rebounds which are prominent in blogs (37\%) of the subset; and \textbf{e.} roughly similar rebounds (6\%).}
    \label{fig:effective_database}
\end{figure*}

The database contains the mentions of approximately 70,000 quotes that appeared in English mainstream media and blogs during a period of 3 months in 2008. Most of the quotes received very few mentions and survived for 2 or 3 days both in the media and blogs, as shown in Figure \ref{fig:effective_database}a. In the same figure we also show a relevant subset of approximately 200 quotes that were highly mentioned in both media and blogs, with good signal-to-noise ratio along periods larger than one week  (see Methods for details). These series are the subject of our analyses and were further re-sampled to a one hour resolution and analyzed in a period of 24 days, starting 3 days before the peak value, and smoothed with a 24 hours rolling window (see Methods for details). 
%<

\section*{Results}

We categorized the series upon visual inspection into four distinct families, as illustrated in Figure \ref{fig:effective_database}b. The quote `{\em the united states is leading diplomatic efforts to achieve a meaningful cease-fire that is fully respected}' is an example of exponential decay in both media and blogs. This behaviour constitutes around the 36\% of the time series in our selected subset.
The remaining time series exhibit more complex dynamics, characterized by damped rebounds of information. For instance, the quote `{\em offered up to a million dollars to name him to the US senate federal law enforcement}' demonstrates a larger amplitude rebound in media compared to blogs (about 21\% of our subset share this behaviour). Conversely, the quote `{\em he was bum-rushed by 200 people}' exhibits the inverse trend (about 37\% of the subset). Additionally, there are instances like `{\em we have sort of become a nation of whiners}' where the magnitude of fluctuations remains consistent across both media and blogs (observed in around the 6\% of the selected subset).

\begin{figure}[htpb]
    \centering
    \includegraphics[width = 7cm]{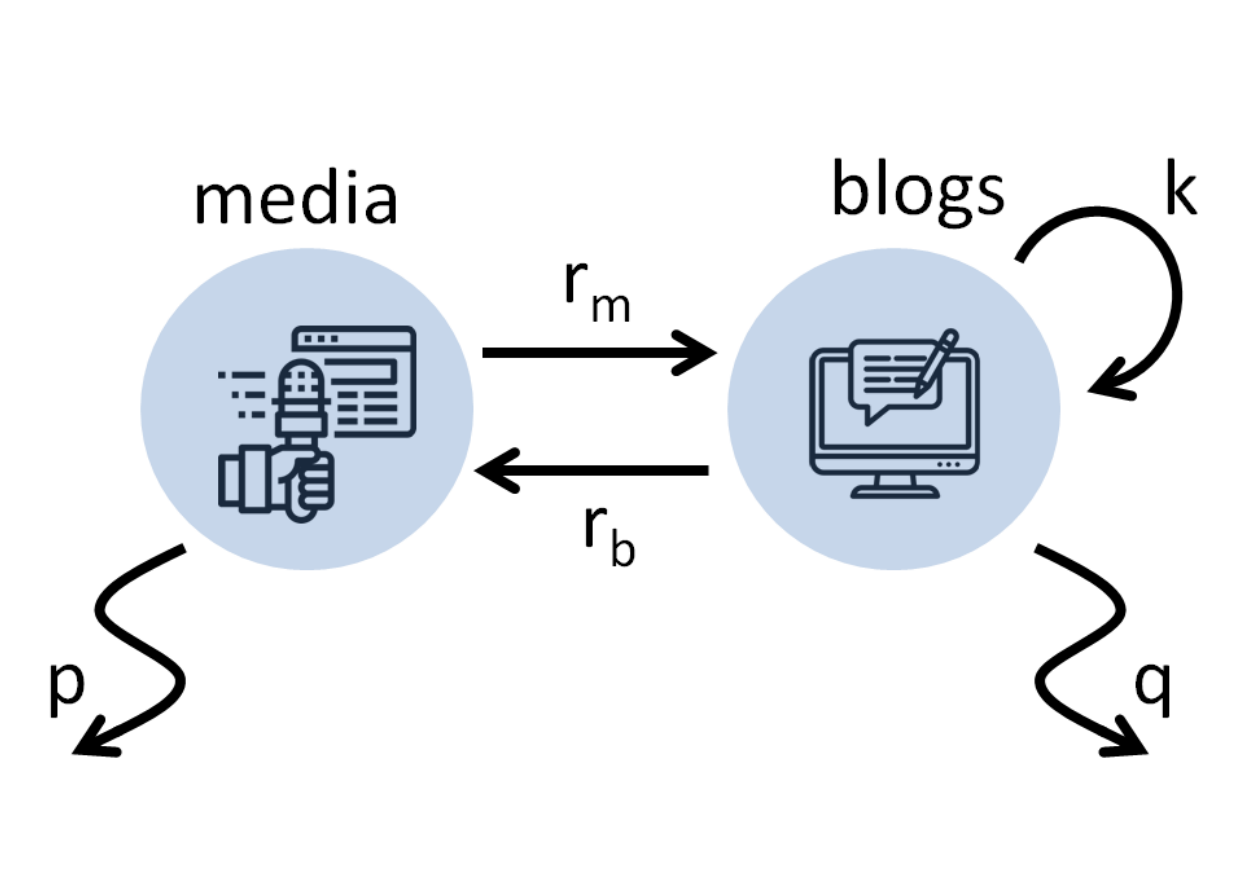}
    \caption{\textbf{Two-way feedback between media and blogs.} On a press release, information flows from the media to the social media at a rate $r_m$. The reactions over a time $\tau$ produce a self-feedback effect $k$ and also feed back to the media at a rate $r_b$. Eventually, collective attention fades at rates $p$ in the media and $q$ in the blogs. This flow of information is made operational in Eq.\ref{eq:1} ($p = 1$ in the dimensionless form). }
    \label{fig:schema}
\end{figure}

To account for these experimental patterns, we present a minimal model of interaction between media and blogs with interpretable parameters, as schematized in Figure \ref{fig:schema}. We assume that the attention in the media and blogs decay at rates $p$ and $q$ and that information flows from media to blogs and back at rates $r_m$ and $r_b$ respectively. This interaction is formalized as follows: 

\begin{align}
    \frac{dm}{dt}   &= -m(t) + r_b\,b(t - \tau) \nonumber \\ 
    \frac{db}{dt}  &= -q\,b(t) + r_m\,m(t) +k\,b(t)\,b(t-\tau).
    \label{eq:1}
\end{align}

The variables $m$ and $b$ represent the volume of information units in the media and blogs respectively. The equations \ref{eq:1} are in dimensionless form, and therefore the decay in blogs is expressed in units of the decay in media (the same holds for the other variables, see Methods for details). 

When $\tau=0$ and $k=0$ we recover the simplest possible model for the feedback process, one in which both variables $b$ and $m$ interact linearly. The general solutions of a two-dimensional linear system include exponential decays and oscillations around the origin \cite{WigginsBook}. Since our variables represent the volume of information units, which are definite positive, oscillations around the origin are forbidden and this system can only explain the observed exponential decays.

To account for the rebounds observed in the other families of experimental data, we assume that information flows much more rapidly from media to blogs than in the opposite direction, which we characterize by a typical time delay $\tau$. Although this delay serves to explain the presence of information rebounds observed experimentally, the relative intensities of these rebounds can be very different in media and blogs (see Figure \ref{fig:effective_database}b). We assume that this results from a self-feedback effect in the information received by people. The form of this non-linear term, $b(t)\,b(t-\tau)$ is inspired in the competition for finite resources in population dynamics of animal species \cite{Lahcen2006}. It states that the public's attention is influenced not only by the volume of mentions at present time, but also by the past mentions. In this approach, we let the control parameter $k$ to be either positive or negative, allowing for an amplification ($k>0$) or inhibition ($k<0$) of the feedback effects.

By construction, this model presents a set of minimum requirements needed to explain the observed dynamics. The result is a two-dimensional linear system with delay and a saturation term, described by 5 independent parameters: the decay $q$ of attention in the public relative to that of media, the rates information flowing from media to blogs $r_m$ and back $r_b$, the self-feedback $k$ and the delay $\tau$.

\begin{figure*}[htpb]
     \centering
    \includegraphics[width = 0.9\textwidth]{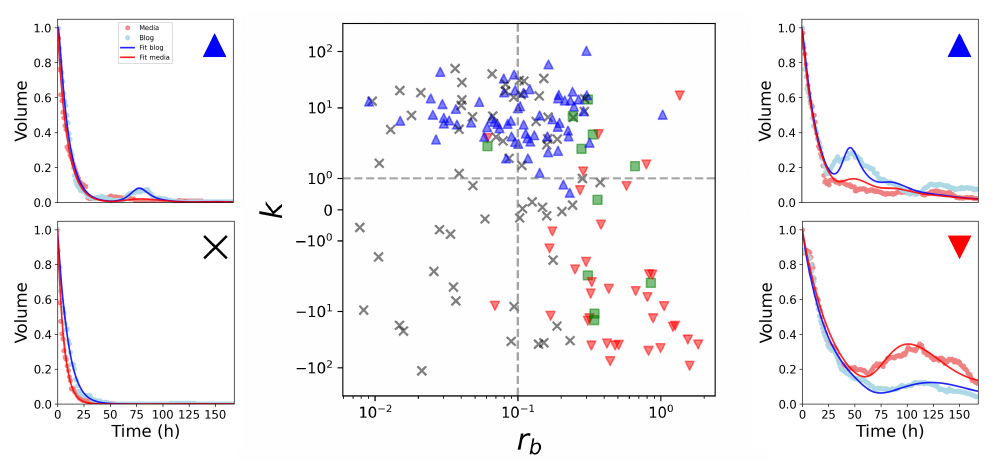}
     \caption{{\bf Data fitting and organization of patterns in the parameter space.} The central panel shows the parameters of the time series fitted with the model in the plane ($r_b,k$). Crosses represent exponential decays. Triangles pointing up show series where rebounds in blogs exceed that of media, and the opposite are marked with triangles pointing down. Similar rebounds in both platforms are marked with squares. Side panels show examples of series of the different regions of the parameter space.} 
     \label{fig:parameters_space_series}
\end{figure*}

We fitted the model to the experimental series using a genetic algorithm (see Methods). Figure \ref{fig:parameters_space_series} shows examples of selected time series fitted by the model and the distribution of solutions in the projection ($r_b,k$) of the parameter space. In this particular projection it becomes apparent that exponential decays emerge when both $r_b$ and $k$ are small. In this case, the equation for media is decoupled from blogs ($dm/dt \sim -m$) and both variables simply decay exponentially, as shown in the lower left panel of Figure \ref{fig:parameters_space_series}.
Information rebounds become more noticeable as the feedback from blogs $r_b$ increases, and the line $k=0$ separates the ones that are higher in blogs (blue dots) than those higher in media (red dots).

Increasing the feedback from blogs is not the only mechanism that produces information rebounds. In fact, when $k>0$ blogs are also affected by the delay, independently from the value of $r_b$. In this case, rebounds are not generated by the forcing of the media, but are self-induced by the interest of blogs themselves. Interestingly, we found examples of this behavior in the experimental data (upper left panel of Figure \ref{fig:parameters_space_series}). 

\begin{figure}[htbp]
     \centering
     \includegraphics[width = \columnwidth]{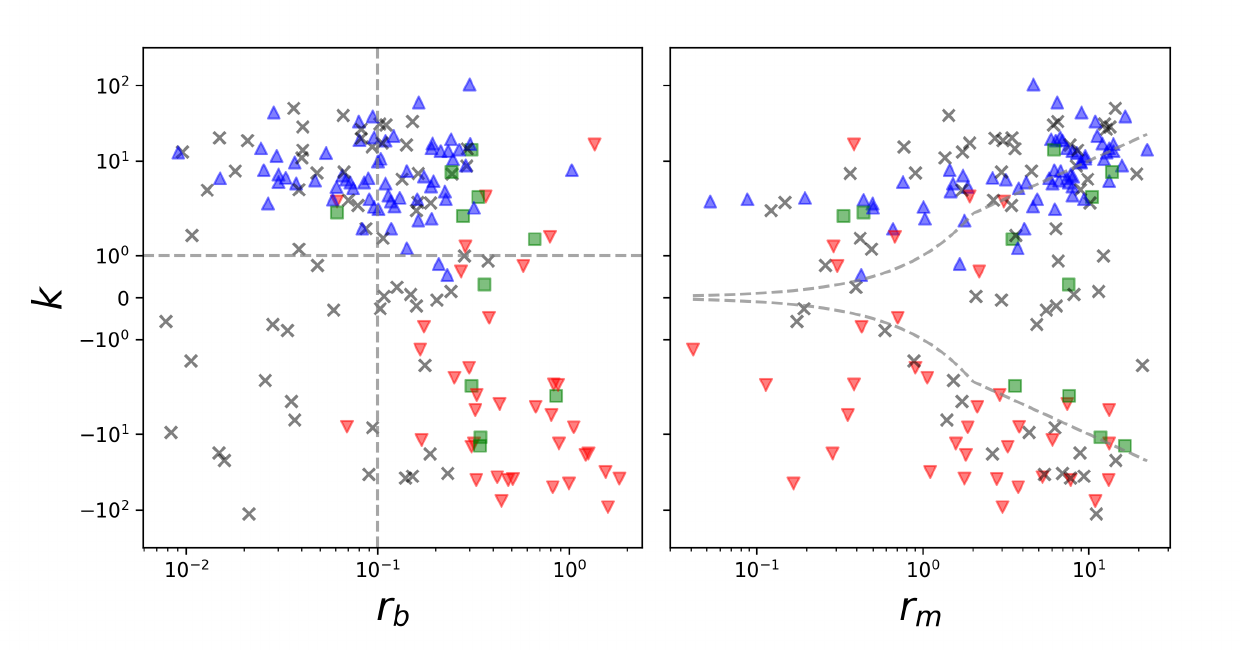}
     \caption{ {\bf Self-feedback and transfer rates between media and blogs}. Projections of the  parameter spaces ($r_b,k$) and ($r_m,k$). Dashed lines in right panel correspond to $|k| = r_m$. Color code is the same as for Figure \ref{fig:parameters_space_series}. 
     }
     \label{fig:parameters_space_k}
\end{figure}

The projections $(r_b,k)$ and $(r_m,k)$ shown in Figure \ref{fig:parameters_space_k} highlight that the information rebounds prevail in the media or in blogs when the self-feedback effect is larger than the forcing of the media, $|k| > r_m$. In the opposite case, when $|k| < r_m$, blogs and media display essentially the same decaying dynamics.

\begin{figure}[htbp]
     \centering
     \includegraphics[width = \columnwidth]{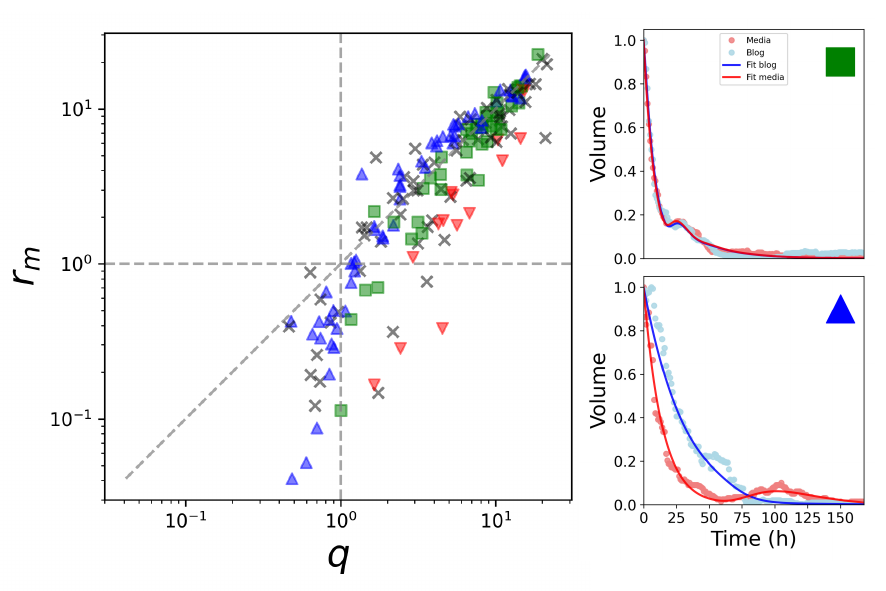}
    \caption{\textbf{Reduction of free parameters.} For $q>1$ and $r_m>1$, data accumulates along the line $q=r_m$. Colors shows different behaviours at the beginning of the series. Blue refers to a faster decay of media respect to blogs, red to a faster decay of blogs respect to media and green roughly equal decay rate.
     Side panels show examples of series of the different regions of the parameter space.}
     \label{fig:r_q}
\end{figure}

Finally, the projection shown in Figure \ref{fig:r_q} discloses an important dynamical property of the system, as data tends to cluster around the line $r_m=q$. This means that units of information that flow fast from media to blogs also decay fast. This finding allows a reduction in the number of free parameters of the system from five to four. In particular, when $|k|$ is small and $q$ is large, blogs passively follow the media (top right panel of Figure \ref{fig:r_q}).

\begin{figure}[htbp]
     \centering
     \includegraphics[width = 6cm]{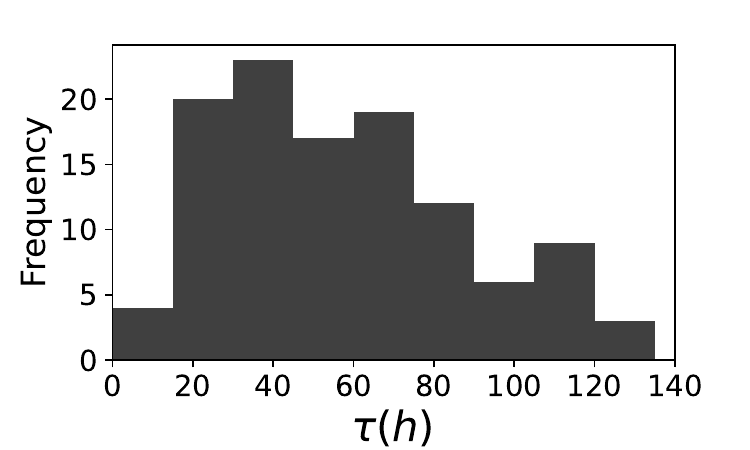}
     \caption{{\bf Distribution of the delay times $\tau$.} The model allows disclosing the typical feedback time of public opinions. Here $\tau$ is expressed in hours, computed from the dimensionless parameter $\tau$ of Eq.\ref{eq:1}).}
     \label{fig:distribution_tau_tilde}
\end{figure}

\begin{figure}[htbp]
     \centering
     \includegraphics[width = \columnwidth]{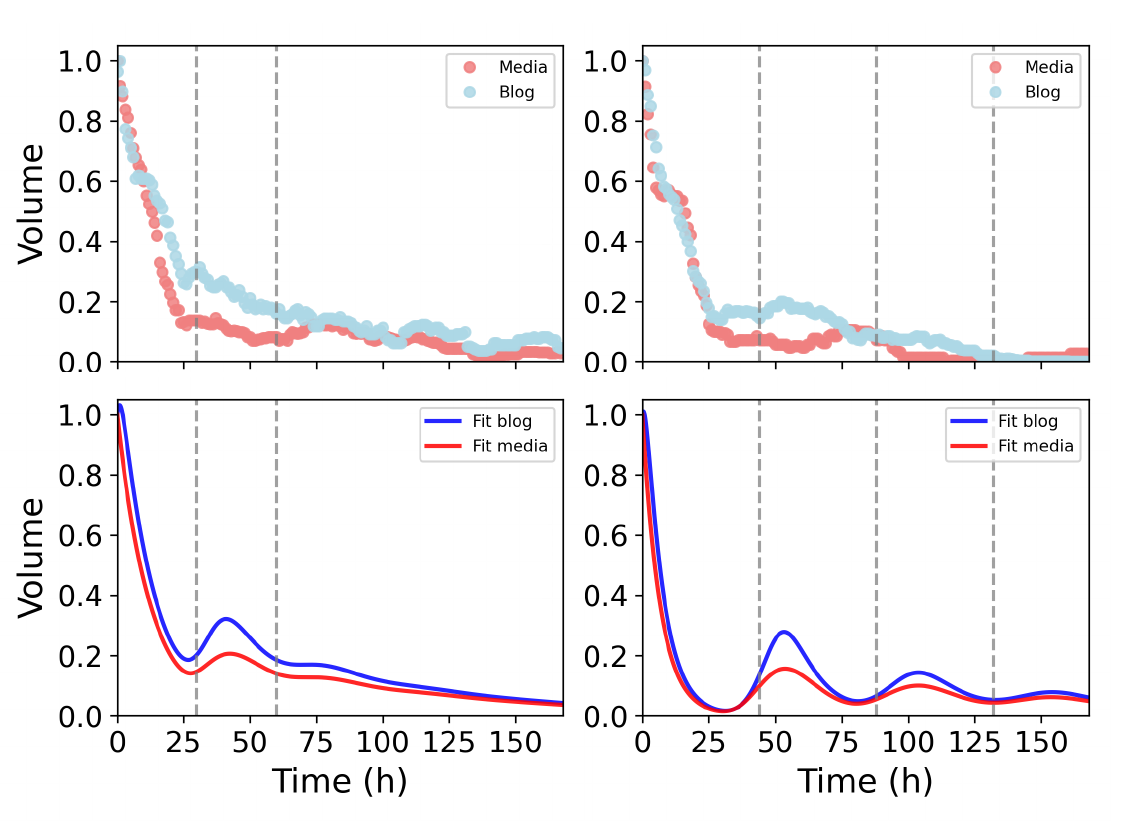}
     \caption{\textbf{Time series with multiple rebounds.} Our model predicts rebounds at multiples of $\tau$. Although rebounds are usually rapidly damped, multiple rebounds can be observed in some series, such as the ones in the upper left panel at $\tau\sim 30$ h and $2\tau$, and the upper right panel at $\tau\sim 44$ h, $2\tau$ and $3\tau$. Bottom panels show qualitative fittings. }
     \label{fig:more_bumps}
\end{figure}

This first set of analyses supports the pertinence of a delay mechanism for explaining the observed patterns of information exchange. Under this hypothesis, data fitting shows that about $80\%$ of the delays $\tau$ occur between $1$ and $4$ days after receiving the information, as shown in Figure \ref{fig:distribution_tau_tilde}, while the mean attention decay is $1/p\sim 13$ hours in media and $1/q\sim 2$ hours in blogs, comparable to the attention decay for hashtags  reported in \cite{lorenz2019accelerating} and showing a faster reaction of blogs when compared to media. It is important to note that a signature of these delay models is the presence of rebounds at multiples of $\tau$. Although in most of the observed time series the effect is too weak, this signature is indeed visible in some experimental series, as the ones shown in the upper panels of Figure \ref{fig:more_bumps} with two and three visible rebounds at multiples of $\tau\sim 30$ h and $\tau\sim 44$ h respectively. Lower panels show simulations with rebounds at similar times, although the relative heights are not accurately reproduced with our minimal model. 

We next investigated the relationship between the observed dynamical patterns and the semantic content of the quotes. For that sake, we first discarded effects due to the length of the quotes or the number of mentioned variants, and also effects due to the total volume of the quotes. We then performed a direct inspection of the content, which we complemented with a summary retrieved by ChatGPT (gpt3.5-turbo), which use a Large Language Model to analyze the content and the tone of each meme.

The series characterized by rebounds in the blogs driven by self-feedback (left upper panel of Figure \ref{fig:parameters_space_series}) predominantly fall within the realm of gossip and entertainment, and tend to employ a more casual language, reflecting individual perspectives. This pattern reflects the fact that this content often trigger discussions on specialized or niche blogs. 
The second group comprises time series with rebounds that are larger in blogs compared to those of the media (upper right panel of Figure \ref{fig:parameters_space_series}). This group shows references to Obama's campaign and victory speeches. The news articles within this group are characterized by high emotional content with a significant impact on debates at both the media and blogs. 
The series that show larger rebounds in media than blogs (lower right panel of Figure \ref{fig:parameters_space_series}) correspond to quotes that tackle harder-hitting topics, primarily revolving around economics and politics such as the acceptance speech of Obama's candidacy. This group exhibits a more critical and sarcastic tone. 
Finally, the group of exponential decays (lower left panel of Figure \ref{fig:parameters_space_series})  includes a diverse range of topics.

\section*{Conclusions}

In this work, we re-analyzed data from a comprehensive database of quotes in the digital environment, uncovering persistent temporal patterns that signify a complex interplay between information sources and recipients. Our study departs from traditional interpretations by conceptualizing media and blogs as distinct entities within the information exchange ecosystem.

We introduced a minimalist dynamical model that effectively captures these observed patterns, establishing a novel understanding of delayed interactions between media and blogs. Our model not only successfully fits a variety of empirical data but also categorizes diverse dynamics into finite categories associated with distinct regions in the parameter space. This categorization extends beyond mere data fitting, suggesting a deeper connection between the dynamical behavior of quotes and their semantic content. Moreover, our findings suggest a functional relationship between the rate of information flow from media to blogs and the rate of attention decay. This insight implies a potential reduction in the model's complexity, enhancing its applicability and interpretation.

However, while our results are encouraging, they represent a first step in understanding the full spectrum of information dynamics in the digital environment. The generalization of our model remains to be tested across more extensive datasets from various countries and over longer periods. Such extended analysis, which is currently underway, is crucial to ascertain whether the mechanisms we identified are universally applicable or if they exhibit significant variations across different digital landscapes.

In conclusion, our study contributes a new perspective to the understanding of information dynamics in the digital era, offering a model that balances simplicity and depth. The insights gained pave the way for future research, promising to deepen our understanding of how information travels and evolves in an increasingly digital world.

\section*{Methods}

\vspace{.5cm}
\noindent{\bf Data.} In the dataset presented in \cite{leskovec2009meme}, the authors analyze a massive amount of news articles and blogs posts and extract a set of quoted phrases usually used to cite people literally. However, quotes are commonly published with variants, and the authors set up a clustering algorithm to identify the root phrases. We take these root phrases as our starting point to study their time evolution. 

For each time series, we re-sampled to a  1 hour by counting the number of quotes within each interval. We identified the global peak (number of quotes both in mainstream media and blogs) and focused on the interval of three days before and 21 days after this peak, in order to identify the period of largest activity.
Finally, we extracted the trend of this period by performing a rolling window average of 24 h.
In summary, we obtained a set of smooth time series that represent the use of quotes and their textual variants both in mainstream media and blogs.

\vspace{.5cm}
\noindent{\bf Dimensionless model.} The original model reads:
\begin{align}
    \frac{dm}{dt} &= -p\,m(t) + r_b\,b(t - \tilde{\tau}) \nonumber \\ 
    \frac{db}{dt} &= b(t)[-q+ k\,b(t - \tilde{\tau})] + r_m\,m(t)  
    \label{eq:model}
\end{align}
with $m(t = 0) = m_0$ and $b(t = 0) = b_0$, where $m(t)$ stands for the amount of meme cited in media and $b(t)$ for this amount in blogs. 
All parameters are defined positive, except $k$ that can be either positive or negative. Figure \ref{fig:schema} shows a schema of the model.
The dimensionless version of Eq.\ref{eq:model} results from the following rescalings: 
\begin{align*}
    p t &\to t \\
    \frac{m(pt)}{m_0} &\to m(t) \\
    \frac{b(pt)}{b_0} &\to b(t)
\end{align*}
The parameters of the dimensionless model are obtained by setting
$p \to 1$, $(qp) \to q$, 
$(r_bm_0/pb_0) \to r_b$, 
$(r_mb_0/pm_0) \to r_m$, and 
$p\tau \to \tau$.
In this dimensionless form, $m(t = 0) = 1$ and $b(t = 0) = 1$.

\vspace{.5cm}
\noindent{\bf Approximation of the delay term.} Integration of differential equations with delay is not straightforward. Here we adopt the approach of converting the delay equations to ordinary differential equations by increasing the dimensionality of the system \cite{Lahcen2006}. The term with delay can be rewritten as
\begin{equation*}
    b(t-\tau) = \int_0^\infty b(t - s) \delta(s-\tau) ds
\end{equation*}
where $\delta$ is the Dirac's delta. The $\delta$-function can be approximated by a gamma density function of order $c$, $g_a^c(u) = \frac{a^cu^{c-1}e^{-au}}{(c-1)!}$.
The mean value of this distribution is $\mu = c/a$ and the variance $\sigma^2 = c / a^2$. We take $a = c / \tau$ in order to fix the mean value $\mu = \tau$. It can be shown that by fixing $\mu = \tau$, $\sigma^2 \to 0$ for higher orders of the distribution ($c \to \infty$), converging to the $\delta$-function centered on $\tau$.
Therefore, after a simple change of variables, we have

\begin{equation*}
    b(t-\tau) \sim \int_{-\infty}^t b(\nu) g_{c/\tau}^c(t-\nu) d\nu.
\end{equation*}

The approximation improves for larger values of $c$. Also, the gamma distribution has the following properties:
\begin{align*}
    \frac{d}{du}g_a^1(u) &= -a g_a^1(u) \\
    \frac{d}{du}g_a^2(u) &= -a (g_a^1(u) - g_a^2(u)) \\
    &... \\
    \frac{d}{du}g_a^j(u) &= -a (g_a^{j-1}(u) - g_a^j(u))
\end{align*}

\noindent with $g_a^1(0) = a$ and $g_a^j(0) = 0$ for $j = 2,...,c$. By defining the following auxiliary variable

\begin{equation*}
    y_j(t) = \int_{-\infty}^t b(\nu) g_{j/\tau}^j(t-\nu) d\nu, 
\end{equation*}

\noindent Eq.\ref{eq:1} can be rewritten as a system of ODEs as:

\begin{align*}
    \frac{dm}{dt}   &= -m(t) + r_b\,y_c(t) \\ 
    \frac{db}{dt} &= b(t)[-q+k\,y_c(t)] + r_m\,m(t) \\
    \frac{dy_1}{dt}   &= \frac{c}{\tau} (b(t)- y_1(t)) \\ 
    &... \\
    \frac{dy_j}{dt}   &= \frac{c}{\tau} (y_{j-1}(t)- y_j(t))
\end{align*}

\noindent with $y_1(0),\,y_j(0) = 0$, for $j = 2,...,c$. The reasoning behind this approximation is that auxiliary variables retard the effect of instant changes in $b$, simulating the presence of the delay in the interaction.

\vspace{.5cm}
\noindent{\bf Fitting of the model.} We apply a genetic algorithm through the python package \emph{scipy} \cite{2020SciPy-NMeth} to achieve the minimum error between the integration of the model and the data. We integrate the model of ODEs obtained after the approximation of the delay term with $c = 30$. No significant changes were observed by varying this number around the same magnitude order.
The error is the define as the mean quadratic distance between model and data in the two dimensional space of media and blogs.

\vspace{.5cm}
\noindent{\bf Data and codes.} Data and codes are available at \url{https://github.com/spinto88/Meme_project}.

\begin{figure}
\centering
\includegraphics[width=0.9\columnwidth]{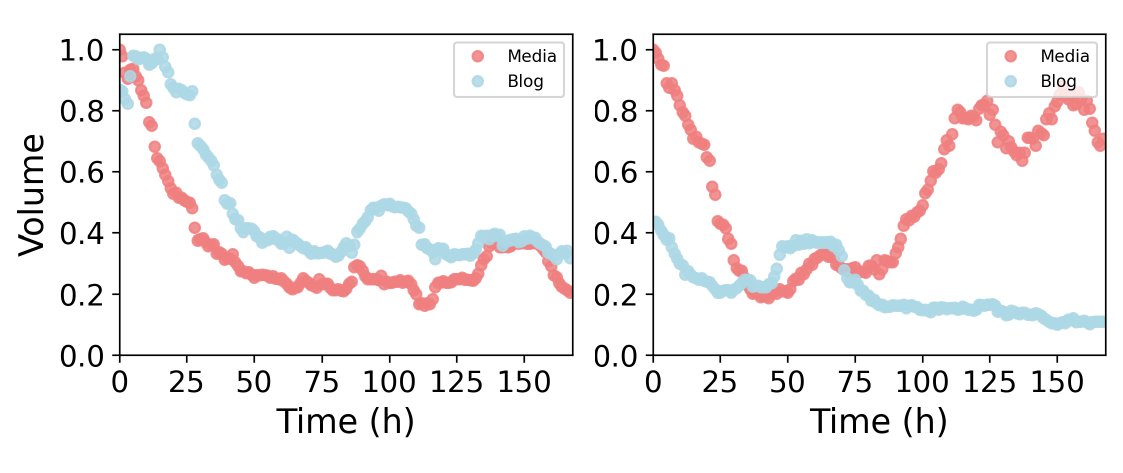}  
\caption{Examples of discarded series.}
\label{fig:discarded_figures}
\end{figure}

\vspace{.5cm}
\noindent{\bf Discarded series.} In Figure \ref{fig:discarded_figures} we show two examples of a small subset of series that we discarded for the analysis. 
We observed cases in which the volume does not decay (left panel) and shows a strong trend (right panel) due to external variables that are not considered in our model.

\vspace{.5cm}
\section*{Acknowledgements}
This research was partially funded by the Universidad de Buenos Aires (UBA)  through Grant UBACyT, 20020220100181BA, the Consejo Nacional de Investigaciones Cientíﬁcas y Técnicas (CONICET) through Grant No. PIP-11220200102083CO, and the Agencia Nacional de Promoción de la Investigación, el Desarrollo Tecnológico y la Innovación through Grant No. PICT-2020-SERIEA-00966.

\bibliography{references} 

\end{document}